\documentclass[twocolumn,preprintnumbers,amsmath,amssymb]{revtex4}

\usepackage{graphicx}
\usepackage{dcolumn}
\usepackage{bm}
\usepackage{epstopdf}

\begin{document}

\title{Connecting Irreversible to Reversible Aggregation: Time and Temperature}

\author{S. Corezzi,$^{1}$ C. De~Michele,$^{2,3}$ E. Zaccarelli,$^{2,3}$ P. Tartaglia,$^{3,4}$ and F. Sciortino$^{2,3}$}
    \affiliation{$^1$CNR-INFM Polylab, Universit$\grave{a}$ di Pisa, Largo
Pontercorvo 3, I-56127 Pisa, Italy\\
$^2$CNR-INFM SOFT, Universit$\grave{a}$ di Roma ``La
Sapienza", Piazzale A. Moro 2, I-00185 Roma, Italy\\
$^3$Dipartimento di Fisica, Universit$\grave{a}$ di Roma ``La
Sapienza", Piazzale A. Moro 2,
I-00185 Roma, Italy\\
$^4$CNR-INFM SMC, Universit\`a di Roma ``La Sapienza", Piazzale A.
Moro 2, I-00185 Roma, Italy}

\date{\today}

\begin{abstract}
We report molecular dynamics simulations of a gel-forming mixture of
ellipsoidal patchy particles with different functionality. We show
that in this model, which disfavors the formation of bond-loops,
elapsed time during irreversible aggregation --- leading to the
formation of an extended network --- can be formally correlated with
equilibrium temperature in reversible aggregation. We also show that
it is possible to develop a parameter-free description of the
self-assembly kinetics, bringing reversible and irreversible
aggregation of loopless branched systems to the same level of
understanding as equilibrium polymerization.
\end{abstract}


\maketitle

Several natural and synthetic materials, as well as biological
structures, result from the self-assembly of elementary units into
branched aggregates and networks \cite{lehn2, supra, leiblernature,
Starr2008, Odriozola}. This self-assembly process is receiving
considerable attention in two fast-growing fields: supramolecular
chemistry \cite{lehn2, supra, leiblernature} and collective behavior
of patchy and functionalized particles \cite{Manoh_03,
MIRKIN_96,23}, among the most promising building blocks of new
materials \cite{Blaad06, Glotz_Solomon_natmat}. The process of
formation of an extended three-dimensional network of bonds
connecting independent molecules, proteins or colloidal particles,
is named gelation and the resulting material a gel \cite{colby,
depablo3,kumar}. The ratio between the bond energy $u_0$ and the
thermal energy $k_BT$ (where $k_B$ is the Boltzmann constant and $T$
is the temperature) can be used to classify aggregation into two
broad categories: for strong attraction strength (\emph{chemical}
case \cite{stockmayer, flory, colby, martin}), bond formation is
irreversible and the number of bonds continuously grows with time.
In the case of weak attraction strength (\emph{physical} case
\cite{zaccajpcm}), bonds break and reform while the number of bonds
progressively reaches its equilibrium value. In the latter case, the
final structure of the system can in principle be predicted with
equilibrium statistical mechanics methods.

The value of the ratio $u_0/k_BT$ separates the two classes.
Conceptually, any model of physical aggregation may be turned into a
chemical model by studying its properties following a quench to
$k_BT \ll u_0$. Similarly, applying temperatures comparable to the
bond energy turns an irreversible aggregation model into a physical
one. The idea of a close connection between irreversible and
reversible aggregation is already contained in the early mean-field
theoretical work of Stockmayer \cite{stockmayer}, considering the
Smoluchowski's kinetic equations solved in the limit of absence of
closed bonding loops. In Stockmayer's calculations, at any time $t$
during chemical aggregation, the distribution of clusters of finite
size $k$ ($N_k(t)$) is identical to that found following equilibrium
statistical mechanics prescriptions, i.e., by maximizing the entropy
with the constraint of a fixed number of bonds.
The $N_k$ are commonly referred to as
Flory--Stockmayer (FS) distributions. Later on, Van Dongen and Ernst
\cite{vandongen} confirmed that the FS distributions are also
solutions of the Smoluchowski's equations when bond-breaking
processes are accounted for. According to these theoretical works
--- based on kinetic equations derived in the limit of
reaction-controlled rates (as opposed to diffusion-controlled
\cite{moreau})  --- a system forming progressively larger and larger
loopless branched aggregates evolves in time via a sequence of
states which are identical to the states explored in equilibrium at
appropriate values of $T$. The equality in the fraction of formed
bonds $p$ (the extent of reaction in chemical language) provides the
connection between $t$ during \emph{reversible} or
\emph{irreversible} aggregation and $T$ in \emph{equilibrium}. Van
Dongen and Ernst~\cite{vandongen} also provide an analytic
expression for the $t$-dependence of $p$ following a sudden change
in the external control parameters, offering the first soluble
example of reversible self-assembly of loopless branched structures.
Interestingly, for functionality two (chain assembly) the solution
coincides with the mean-field analytic expressions later on derived
for equilibrium polymerization kinetics~\cite{catescandau}. Despite
their relevance, to our knowledge the Van Dongen and Ernst
predictions have never been tested experimentally or numerically to
assess their limit of validity.

In this Letter --- stimulated by the renewed interest in reversible
and irreversible self-assembly of nano- and micro-particles
interacting via specific directional bonds \cite{lehn2, supra,
Manoh_03, MIRKIN_96, 23} --- we provide a stringent test of the
suggested connections between chemical and physical aggregation, and
establish the limits of analytic description of the kinetics of
formation of branched structures. We do so by extending a model for
chemical gels, recently introduced~\cite{soft}, to the corresponding
physical case by turning the attraction strength between bonding
sites finite. The model, originally inspired by epoxy-amine
step-polymerization \cite{Corezzi2}, represents two types
of mutually reactive molecules $A$ and $B$ as hard homogeneous
ellipsoids of revolution whose surface is decorated in a predefined
geometry by $f_A$ and $f_B$ identical reactive sites. The study of
the chemical version of this model \cite{soft} showed that formation
of closed bonding loops in finite size clusters is disfavored,
possibly due to the non-spherical particle shape and the location of
the reactive sites. Therefore, it offers to us the possibility to
carefully check the Stockmayer \cite{stockmayer} and Van Dongen and
Ernst \cite{vandongen} predictions. We can assess how closely
--- in the absence of loop formation --- the chemical
gelation process in a system of functionalized or patchy units, as
well as the reversible evolution from a monomeric to an equilibrium
bonded state, can be envisaged as a progressive sequence of
equilibrium states, closely connecting $t$ with $T$.

\begin{figure}[b]
\vspace{-0.8 cm}
\includegraphics[width=8.5 cm]{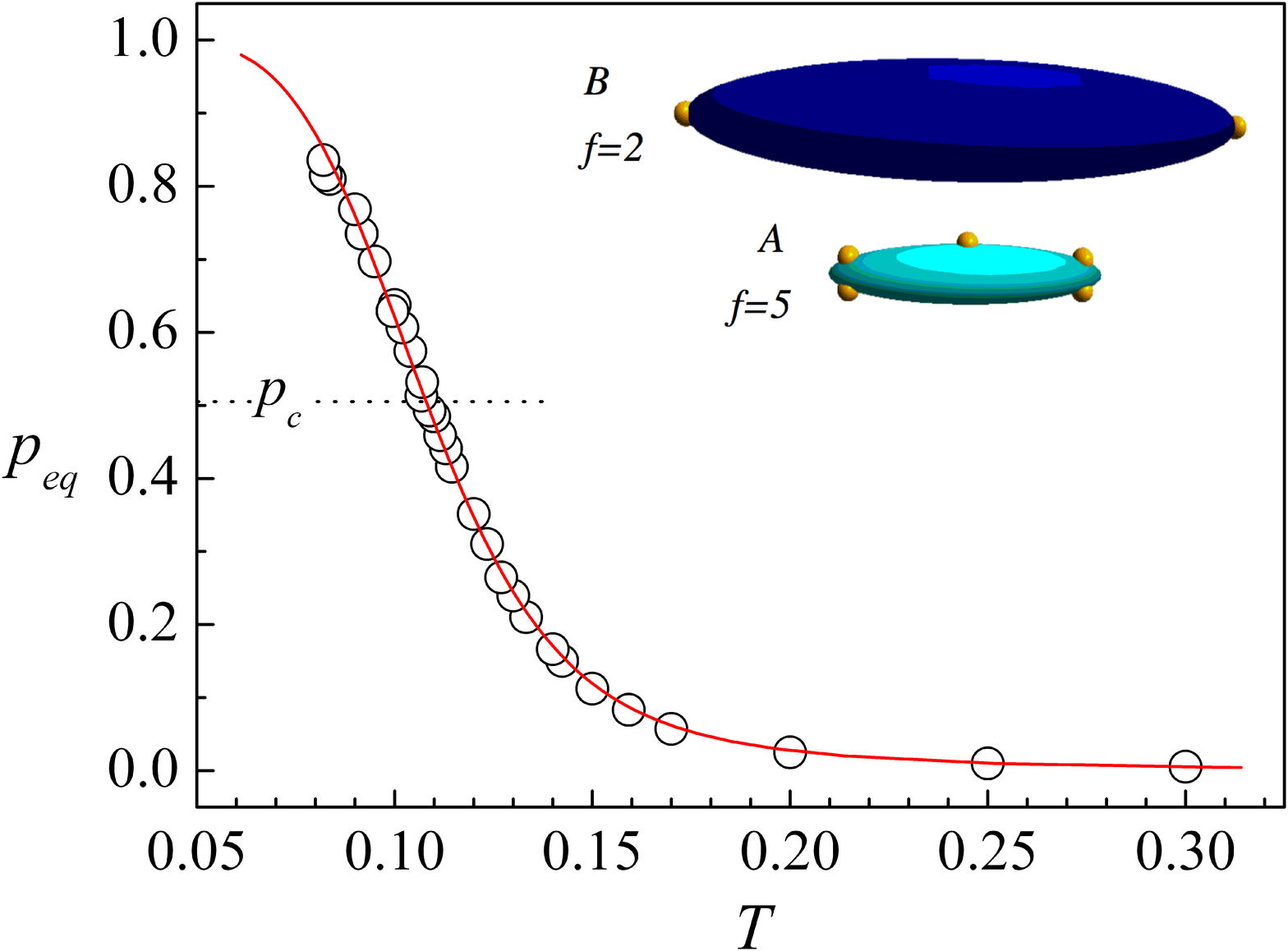}
\vspace{-0.5 cm} \caption{\label{fig:relationPT} Temperature
dependence of the fraction
  of bonds at equilibrium, $p_{eq}$. Symbols: simulation results. The
  solid line represents $p_{eq}(T)$ from Eq.~\ref{eq:Wertheim}.  A
  one-parameter best-fit gives $\Delta S/k_B= 8.52$. The percolation
  threshold $p_{c}=0.5$ is indicated by a horizontal line. The inset shows
  a graphic description of the $A$ and $B$ particles.
  The centers of the small spheres locate the bonding sites on the
  surface of the hard--core ellipsoid.}
\end{figure}

We study via event-driven molecular dynamics simulations a binary
mixture composed of $N_{A}=480$ pentafunctional ($f_A=5$) ellipsoids
of type $A$ and $N_{B}=1200$ bifunctional ($f_B=2$) ellipsoids of
type $B$, so that the number $N_{A}f_{A}$ of $A$-type reactive sites
equals the number $N_{B}f_{B}$ of $B$-type reactive sites.
$A$-particles have mass $m$, revolution axis $a=10\sigma$ and the
other two axes $b=c=2\sigma$; $B$-particles have mass $3.4 m$, and
axes $a=20\sigma$, $b=c=4\sigma$. Size- and mass-ratio are chosen to
mimic the values of an epoxy-amine system \cite{Corezzi2}. The
interaction potential is the hard ellipsoid potential $V_{HE}$
supplemented by site-site square-well attractive interactions
$V_{SW}$ (of strength $u_{0}$, and width $\delta=0.2\sigma$) between
pairs of particles of different type. Our unit mass is $m$, the unit
energy $u_{0}$. $T$ is measured in units of the potential depth
(i.e., $k_{B}=1$), time $t$ in units of $\sigma (m/u_{0})^{1/2}$.
The packing fraction is fixed at $\phi=0.3$ and $T$ is varied from
$T=0.3$ to $T=0.065$. Event-driven (newtonian) molecular dynamics
simulations (in the NVT ensemble) are performed using a specifically
designed code \cite{code}. Two sites, on particles of different
type, form a bond if their distance is closer than $\delta$.
Clusters are defined as groups of bonded particles. The particle
shape and the location of the attractive sites (see inset of
Fig.~\ref{fig:relationPT}) is such that each site is engaged at most
in one bond, ensuring an unambiguous definition of the extent of
reaction $p$ by the ratio between the number of bonds in the system
and the maximum number of possible bonds (i.e., $N_{A}f_{A}$).
Unless otherwise stated, we study the trajectory of the system in
configuration space, starting from an initial configuration with no
bonds between particles, at fixed $T$. During the simulation, bonds
form and break continuously with time while the system evolves
toward the equilibrium state, characterized by an equilibrium value
$p_{eq}(T)$ of the extent of reaction. To improve the statistics we
average over 11 different realizations for each studied $T$.
Equilibrium results are obtained from the final part of the
trajectories, after the equilibration transient is over.

Figure~\ref{fig:relationPT} shows $p_{eq}(T)$, for all $T$ values
where equilibration was achieved within the allocated computational
time (six months at low $T$). Within the range $0.2 > k_BT/u_0 >
0.08$ the system crosses from a monomeric state to a significantly
bonded state, with more than 80\% of the bonds formed. The sharp
sigmoidal shape can be perfectly represented by the independent-bond
mass-action law \cite{Werth1}
\begin{equation}
\frac{p_{eq}}{(1-p_{eq})^2} = e^{\beta (u_0 - T \Delta S)}
\label{eq:Wertheim}
\end{equation}
where $u_0$ and $\Delta S$ describe, respectively, the energy and
the entropy change associated to the formation of a single bond.

\begin{figure}
\includegraphics[width=9.0 cm]{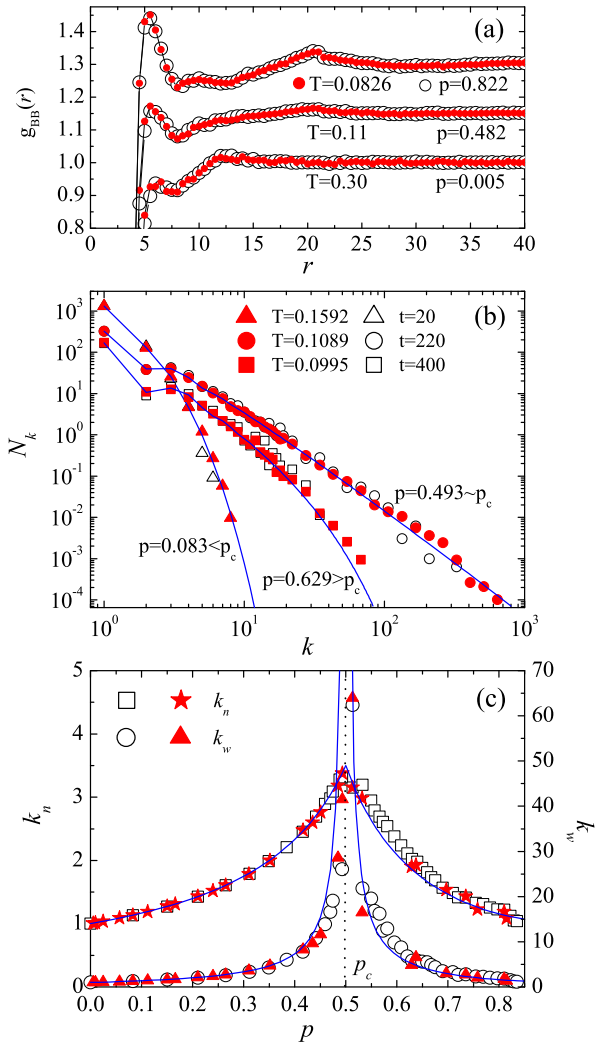}
\vspace{-0.6 cm} \caption{\label{fig:DISTR} (a) Partial radial
distribution function
  for $B$ particles, $g_{BB}(r)$; (b) cluster size distribution $N_{k}$ and
  (c) number- and weight-average size of the finite
  clusters ($k_{n}$ and $k_{w}$). In all panels, equilibrium results
  (closed symbols) are compared with results at different times $t$
  during irreversible aggregation (open symbols, from
  Ref.~\onlinecite{soft}) such that $p(t)=p_{eq}(T)$. In panel (a),
  curves at T=0.11 and T=0.0826 are shifted by +0.15 and +0.30, respectively.
  In (b) and (c) solid lines indicate FS predictions.}
\end{figure}

Figures~\ref{fig:DISTR}(a)-(b) show the equilibrium structure of the
system and the cluster size distribution $N_k$ at three different
values of $T$, corresponding to $p_{eq}$ values below, at and above
percolation (numerically located at $p_{c}=0.505\pm 0.007$
\cite{soft}). To describe the structure of the system we report the
center-to-center pair distribution function $g_{BB}(r)$ for the $B$
particles. To quantify the $T$ dependence of $N_k$ we also report
the number- ($k_n$) and weight-average size ($k_w$) of the finite
clusters (Fig.~\ref{fig:DISTR}(c)). Data shown in
Figs.~\ref{fig:DISTR}(b)-(c) are very well described
---without any fitting parameter--- both below and above percolation
by the FS predictions \cite{flory, soft} specialized to the
$f_A$=5--$f_B$=2 case, confirming that bonding loops in finite size
clusters can be neglected. Therefore, the model allows us to check,
without fitting parameters, if the evolution of the system during
equilibration and irreversible aggregation does follow a sequence of
equilibrium steps, and in particular, how $t$ in chemical gelation
can be associated to a corresponding equilibrium $T$. To this aim,
we compare in Figs.~\ref{fig:DISTR}(a)-(c) the equilibrium
quantities with the corresponding quantities evaluated at selected
elapsed times $t$ during the irreversible bonding process. The
specific $t$ value is chosen in such a way that $p(t)=p_{eq}(T)$.
For all these quantities, at each $t$, data are identical to those
obtained in equilibrium at the same extent of reaction,
demonstrating that the evolution of the system structure and
connectivity during irreversible aggregation does follow a sequence
of equilibrium states.

\begin{figure}
\includegraphics[width=9 cm]{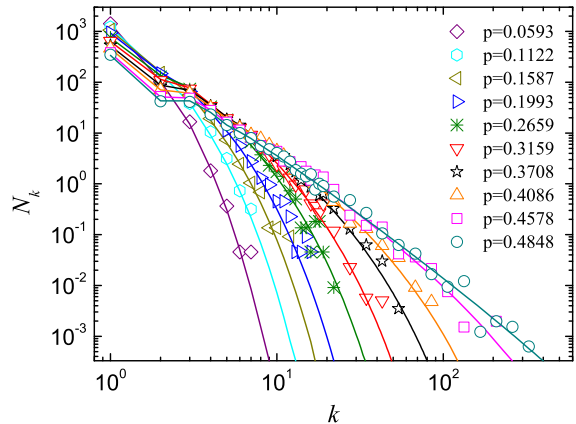}
\vspace{-0.7 cm} \caption{\label{fig:T0e11} Distribution of finite
size clusters $N_{k}$ for different times (and fraction of bonds
$p$) during reversible aggregation following a $T$-jump to $T=0.11$.
At each time during the equilibration process, the cluster size
distribution is identical to the FS distribution at the same
fraction of bonds (solid lines)}.
\end{figure}

Also the equilibration process at finite $T$, i.e., when
bond-breaking processes are included, proceeds along a sequence of
equilibrium states described by the FS theory. Indeed,
Fig.~\ref{fig:T0e11} shows $N_{k}(t)$ for several successive times
following a quench, in conditions of reversible aggregation, from a
high-$T$ unassembled state to a low-$T$ assembled one. Data are
compared --- without fit parameters --- with the corresponding
equilibrium quantities, accurately modeled by the FS distributions.
Similar agreement is observed for all the investigated equilibration
processes.

\begin{figure}
\vspace{-0.8 cm}
\includegraphics[width=9 cm]{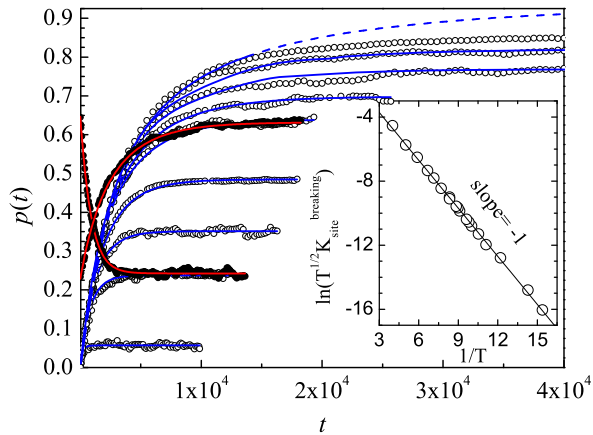}\vspace{-0.5 cm}
\caption{\label{fig:PvsTIME} Time dependence of the fraction of bonds
  $p(t)$ during reversible aggregation at different temperatures
  ($T$=0.17, 0.13, 0.12, 0.11, 0.10, 0.095, 0.09, 0.082, 0.07 from the
  lower to the higher equilibrium value $p(\infty)=p_{eq}$), starting
  from an unbonded configuration ($p(0)=0$; open symbols) or from a generic equilibrated
  configuration ($p(0) \ne 0$; closed symbols).
  Lines are solutions of the Smoluchowski's equation for reversible aggregation
  (Eq.~\ref{eq:ptf}), with $p_{eq}$ from Eq.~\ref{eq:Wertheim} and
  $K_{site}^{breaking}$ as the only fit-parameter.
  The inset shows the $T$ dependence of $K_{site}^{breaking}$.}
\end{figure}

Data in Fig.~\ref{fig:T0e11} show that the cluster size distribution
during the kinetics of equilibration evolves via a sequence of FS
distributions. Fig.~\ref{fig:PvsTIME} shows the associated time
dependence of the bond probability $p(t)$, following a $T$-jump
starting either from a high-$T$ unbonded configuration ($p(0)=0$) or
from a previously equilibrated configuration characterized by
branched aggregates ($p(0) \ne 0$). Lines in Fig.~\ref{fig:PvsTIME}
are the theoretical predictions \cite{vandongen} ---with the Flory
post--gel assumption \cite{flory} for values of $p(t)$ above
percolation--- i.e.
\begin{equation}
p(t)=p_{eq} \frac{1-\left( \frac{1-p(0)/p_{eq}}{1-p(0)p_{eq}}
\right)\rm{e}^{-\Gamma t}}  {1-p_{eq}^{2}\left(
\frac{1-p(0)/p_{eq}}{1-p(0)p_{eq}} \right)\rm{e}^{-\Gamma t}}~,
\label{eq:ptf}
\end{equation}
with $\Gamma \equiv K_{site}^{breaking} (1+p_{eq})/(1-p_{eq})$,
where $K_{site}^{breaking}$ is the rate constant for breakage of a
single bond. Since $p_{eq}$ is known from Eq.~\ref{eq:Wertheim}, the
entire equilibration dynamics in this aggregating branching system
only depends on one parameter, $K_{site}^{breaking}$, which fixes
the time scale of the aggregation process. The theoretical
expression very well represents the numerical data, except for the
two lowest studied temperatures ($T\leq 0.07$), at which extensively
bonded states are reached ($p>0.8$). Such disagreement can be traced
to a failure of the Flory post-gel assumption and/or to a
progressive role of the size-dependence of the cluster mobility. The
overall agreement between theory and simulation in
Fig.~\ref{fig:PvsTIME} is further stressed by the $T$ dependence of
the single fit-parameter $ K_{site}^{breaking} \sim
e^{-u_0/k_BT}/\sqrt{T}$ (see inset), which as expected incorporates
both an Arrhenius term and the thermal velocity component $\sim
\sqrt{T}$, entering the attempt rate of bond breaking.

The results of this study clearly indicate that the irreversible
evolution of a system of patchy or functionalized particles, in
which bonding loops can be neglected, can be put in correspondence
with a sequence of equilibrium states. For this class of aggregating
systems, it is thus possible to convert irreversible-aggregation
time into an effective $T$ and to envisage the evolution of a
chemical gel as a progressive cooling of the corresponding physical
model. Elapsed time $t$ can be uniquely associated to an equilibrium
$T$, recalling the concept of fictive $T$ in aging glasses
\cite{tfictive}. Equally important is the possibility of
interpreting cases in which during the formation of a chemical gel
the corresponding thermodynamic path crosses a thermodynamic
instability line, e.g., the gas--liquid coexistence line \cite{23},
producing an inhomogeneous arrested structure. Such thermodynamic
lines have been recently  calculated with statistical mechanics
methods and computer simulations \cite{bian, 23, Zacca1}. Thus, it
will become possible to interpret the stability and structural
properties of chemical gels by connecting them to the thermodynamic
properties and to the phase diagram of the corresponding physical
models.

Our study also demonstrates, for a realistic model of patchy
particles, that the self-assembly kinetics of particles aggregating
in loopless structures can be fully described ---with no fitting
parameters--- by merging a thermodynamic approach providing
$p_{eq}(T)$ (e.g., Eq.~\ref{eq:Wertheim}) with the Smoluchowski's
equations studied by Van Dongen and Ernst, thus bringing reversible
and irreversible aggregation of loopless branched systems to the
same level of understanding as the equilibrium polymerization case
\cite{catescandau}. This is made possible  by the limited valence of
the particles and by the specificity of the site-site interaction.
The reduced valence, coupled to the non-spherical particle shape,
contributes to disfavor the formation of closed bond-loops, while
the specificity in the bonding interaction contributes to reduce the
rate-controlling role of diffusion compared to the bonding process
\cite{moreau}.

We acknowledge support from NoE SoftComp NMP3-CT-2004-502235. We
thank D. Fioretto and P.A. Rolla for helpful discussions.

\bibliography{biblio}

\end{document}